\documentclass[prb,superscriptaddress,twocolumn]{revtex4-1}
\usepackage[dvips]{graphicx}
\usepackage{subfigure}
\usepackage{bm}
\usepackage{color}

\begin{document}

\title{
Spinmotive force with static and uniform magnetization \\induced by a time-varying electric field
}

\author{Yuta Yamane}
\affiliation{Advanced Science Research Center, Japan Atomic Energy Agency, Tokai,
Ibaraki 319-1195, Japan}
\author{Jun'ichi Ieda}
\affiliation{Advanced Science Research Center, Japan Atomic Energy Agency, Tokai,
Ibaraki 319-1195, Japan}
\affiliation{CREST, Japan Science and Technology Agency, Tokyo 102-0075, Japan\\}
\author{Sadamichi Maekawa}
\affiliation{Advanced Science Research Center, Japan Atomic Energy Agency, Tokai,
Ibaraki 319-1195, Japan}
\affiliation{CREST, Japan Science and Technology Agency, Tokyo 102-0075, Japan\\}

\date{\today}


\begin{abstract}
A new spinmotive force is predicted in ferromagnets with spin-orbit coupling.
By extending the theory of spinmotive force, we show that a time-varying electric field can induce a spinmotive force with static and uniform magnetization.
This spinmotive has two advantages;
it can be detected free from the inductive voltage owing to the absence of dynamical magnetization and it can be tuned by electric fields.
To observe the effect, we propose two experimental setups:
electric voltage measurement in a single ferromagnet and spin injection from a ferromagnet into an attached nonmagnetic conductor.
\end{abstract}

\maketitle


\section{Introduction}
Investigation of mutual interaction between electrons and magnetization is a key subject in the field of spintronics.\cite{spincurrent}
Spinmotive force (SMF) is an emerging concept,\cite{barnes} where spin current and electric voltage are induced in a ferromagnet due to the exchange coupling between electrons and the magnetization.
The SMF provides an important ground for the basic study of the electron-magnetization interaction,\cite{barnes,volovik,duine,tserkov,yamane,shibata,sanvito,duine2,sfzhang2,duine3,pma,yang,hayashi,sfzhang,shibata2,niu,ohe,tanabe,comb,lee,tatara} as well as a new concept for spintronic devices.\cite{barnes2,hai,ieda}
The SMF is described by the so-called spin electric field, which accelerates electrons in opposite directions depending on their spin, giving rise to a spin current in the ferromagnet and then an electric voltage.
Until recently, adiabatic and nonadiabatic contributions to the spin electric field have been derived, which depend on both $\partial{\bm m}/\partial t$ and $\nabla{\bm m}$,\cite{barnes,volovik,duine,yamane,tserkov,shibata,sanvito} where ${\bm m}$ is the classical unit vector of the magnetization direction, and $\partial/\partial t$ and $\nabla$ represent the derivatives with respect to time and space, respectively.
Therefore, the appearance of the SMFs is confined in time-varying and spatially nonuniform magnetization regions, such as moving domain walls,\cite{yang,hayashi} vortex cores,\cite{tanabe} and asymmetrically patterned films.\cite{comb}
Recently, it has been pointed out that in a system with Rashba spin-orbit (SO) coupling there exist additional spin electric fields relying only on $\partial{\bm m}/\partial t$.\cite{lee,tatara}
The discoveries have enabled us to generate a SMF in time-varying but spatially uniform magnetic structures, such as ferromagnetic resonance systems.

In this work the theory of SMF is further extended in a system with SO coupling, in which a sample is subjected to electric fields that can vary in time. 
We predict a new SMF relying on neither $\partial{\bm m}/\partial t$ nor $\nabla{\bm m}$;
a new spin electric field is found to be proportional to ${\bm m}\times\partial{\bm E}/\partial t$, with ${\bm E}$ a U(1) electric field.
Thus the SMF can be generated due to time-varying electric fields with static and uniform magnetization.
This SMF has two advantages compared with the other forms of SMFs.
(i) The electrical measurement of the SMF is free from the inductive voltage because of no dynamical magnetization.
(ii) The SMF can be tuned by the electric fields free from the characteristic frequencies inherent in ferromagnets, such as the ferromagnetic resonance frequency.
We demonstrate the SMF in two systems:
electric voltage measurement in a single ferromagnet and spin injection from a ferromagnet into nonmagnetic conductor.


\section{Formalism}
In a nonrelativistic limit up to the order of $1/c^2$ with $c$ the speed of light, the Hamiltonian of a conduction electron in a ferromagnetic conductor is written by
\begin{equation}
\mathcal{H} = \frac{\bm{p}^2}{2m_{\rm e}} + J_\mathrm{ex} \bm{\sigma}\cdot\bm{m} - \frac{e\eta_\mathrm{so}}{\hbar}\bm{\sigma}\cdot\left(\bm{p} \times \bm{E} \right), 
\label{h}\end{equation}
where $m_{\rm e}$ and $-e$ are the electron's mass and charge.
The second term represents the exchange interaction between the electron spin and the magnetization, with $J_\mathrm{ex}$ the exchange coupling energy and $\bm{\sigma}$ the Pauli's matrices.
In the third term we introduce a SO interaction, with the SO coupling parameter $\eta_\mathrm{so} = \hbar^2/4m_{\rm e}^2c^2$ for the free electron model, but in real materials $\eta_{\rm so}$ can be enhanced by several orders of magnitude.\cite{winkler}
The magnetization ${\bm m}$ and the electric field ${\bm E}$ are in general dependent on time and space.

To derive SMFs, let us investigate the equation of motion for the conduction electron.
The velocity operator ${\bm v}$ is given by ${\bm v}=(1/i\hbar)[{\bm r},{\cal H}] = {\bm p}/m_{\rm e} + (e\eta_{\rm SO}/\hbar){\bm \sigma}\times{\bm E}$, where the second term in the last line is the so-called anomalous velocity.
The ``force" $\bm{\mathcal{F}}$ acting on the electron is given by $\bm{\mathcal{F}} = (1/i\hbar)[m_{\rm e}{\bm v},{\cal H}] + \partial(m_{\rm e}{\bm v})/\partial t$, which is explicitly expressed as
\begin{eqnarray}
\bm{\mathcal{F}} &=& -J_{\rm ex}{\bm \sigma}\cdot\nabla{\bm m} + \frac{em_{\rm e}\eta_{\rm so}}{\hbar}{\bm \sigma}\times\frac{\partial{\bm E}}{\partial t}\nonumber\\
&&+ \frac{em_{\rm e}\eta_{\rm so}J_{\rm ex}}{\hbar}[{\bm \sigma}\times{\bm E},{\bm \sigma}\cdot{\bm m}].
\label{ff}\end{eqnarray}
The first term in the right-hand side of Eq.~(\ref{ff}) is just the spatial gradient of the potential energy $J_{\rm ex}{\bm \sigma}\cdot{\bm m}$, while the second term originates from the time derivative of the anomalous velocity.
The third term is due to the noncommutating nature of the anomalous velocity and the exchange coupling energy.
Equation~(\ref{ff}) is an SU(2) operator containing the Pauli matrices $\bm{\sigma}$ that play as the electron spin operators.

Next, the expectation value of the force $\langle{\bm k}\pm|\bm{{\cal F}}|{\bm k}\pm\rangle$ is calculated by determining the electron spin dynamics $\langle{\bm k}\pm|{\bm \sigma}|{\bm k}\pm\rangle$, where $|\bm{k}\pm\rangle$ stands for a one electron state with momentum $\hbar\bm{k}$ and majority ($+$) or minority ($-$) spin.
It is assumed that the electron spin dynamics is described by a Bloch-type equation of motion,
\begin{equation}
\frac{\partial}{\partial t}\langle{\bm k}\pm|{\bm \sigma}|{\bm k}\pm\rangle = - \frac{2J_{\rm ex}}{\hbar} \langle{\bm k}\pm|{\bm \sigma}|{\bm k}\pm\rangle\times{\bm m} - \frac{\delta{\bm m}_{\pm}}{\tau_{\rm sf}},
\label{eom}
\end{equation}
where $\tau_{\rm sf}$ is the relaxation time for the electron spin flip, and $\delta{\bm m}$ represents a misalignment between the electron spin and the magnetization, which is defined as $\langle{\bm k}\pm|{\bm \sigma}|{\bm k}\pm\rangle= \mp {\bm m} + \delta{\bm m}_\pm$.
The first term in the right-hand side of Eq.~(\ref{eom}) is the Larmor precession around the magnetization ${\bm m}$ caused by the exchange coupling.
On the other hand, the second term arises because of the SO coupling with impurities, other electrons and so on, describing a damping motion towards $-{\bm m}$ ($+{\bm m}$) for the majority (minority) spin.
Equation~(\ref{eom}) implicitly assumes the condition $J_{\rm ex}\gg e\eta_{\rm so}|{\bm k}||{\bm E}|$, where the electron spin dynamics is mostly dominated by ${\bm m}$ due to the relatively strong exchange coupling, while the SO interaction still plays an important role causing the relaxation motion through the nonadiabatic spin flip process.

The misalignment $\delta{\bm m}_\pm$ is essential for $\langle{\bm k}\pm|\bm{{\cal F}}|{\bm k}\pm\rangle$, although in general it is small compared to the component $\mp{\bm m}$.
One can easily see that the values, $\langle{\bm k}\pm|{\bm \sigma}\cdot\nabla{\bm m}|{\bm k}\pm\rangle$ and $\langle{\bm k}\pm|[{\bm \sigma}\times{\bm E}, {\bm \sigma}\cdot{\bm m}]|{\bm k}\pm\rangle$, appearing in the force are zero if $\langle{\bm k}\pm|{\bm \sigma}|{\bm k}\pm\rangle=\mp{\bm m}$.
Without loss of generality, we can decompose $\delta{\bm m}_{\pm}$ into two directions perpendicular to the magnetization as $\delta{\bm m}_\pm= X_\pm {\bm m}\times d{\bm m}/dt+ Y_\pm d{\bm m}/dt$, where $X_\pm$ and $Y_\pm$ are spin-dependent constants, and $d/dt$ is the Lagrange derivative as $d/dt=\partial/\partial t + \langle{\bm k}\pm|{\bm v}|{\bm k}\pm\rangle\cdot\nabla$.
By substituting $\langle{\bm k}\pm|{\bm \sigma}|{\bm k}\pm\rangle$ with this expression of $\delta{\bm m}_\pm$ into Eq.~(\ref{eom}) and comparing the left-hand and right-hand sides, the explicit forms of $X_\pm$ and $Y_\pm$ are determined.
In the process, the term $\partial\delta{\bm m}_\pm/\partial t$, which is the order of $\partial^2{\bm m}/\partial t^2$, is discarded.
The electron spin is in the end expressed in terms of the magnetization as\cite{yamane}
\begin{equation}
\langle{\bm k}\pm|{\bm \sigma}|{\bm k}\pm\rangle = \mp \left[ {\bm m} - \frac{\hbar}{2J_{\rm ex}}\left( {\bm m}\times\frac{d{\bm m}}{dt} + \frac{\hbar}{2J_{\rm ex}\tau_{\rm sf}}\frac{d{\bm m}}{dt}\right) \right].
\label{spin}\end{equation}

Substituting Eq.~(\ref{spin}) into the expectation value of Eq.~(\ref{ff}), we obtain
\begin{equation}
\langle{\bm k}\pm|\bm{{\cal F}}|{\bm k}\pm\rangle = - e \bm{{\cal E}}_\pm,
\label{force}\end{equation}
 where $\bm{{\cal E}}_\pm$ is the spin electric field that is given by 
\begin{eqnarray}
\bm{{\cal E}}_\pm &=& \pm\frac{\hbar}{2e}{\bm m}\times\frac{\partial{\bm m}}{\partial t}\cdot\nabla{\bm m} \pm \frac{\hbar}{2e}\frac{\hbar}{2J_{\rm ex}\tau_{\rm sf}}\frac{\partial{\bm m}}{\partial t}\cdot\nabla{\bm m}\nonumber\\
&& \pm \frac{m_{\rm e}\eta_{\rm so}}{\hbar} \frac{\partial}{\partial t}\left({\bm m}\times{\bm E} \right)\nonumber\\
&& \mp \frac{m_{\rm e}\eta_{\rm so}}{\hbar} \frac{\hbar}{2J_{\rm ex}\tau_{\rm sf}}\left({\bm m}\times\frac{\partial{\bm m}}{\partial t}\right)\times{\bm E}.
\label{es}\end{eqnarray}
In Eq.~(\ref{force}) the velocity-dependent terms are discarded, which include the spin magnetic fields causing the anomalous Hall effect due to the scalar spin chirality\cite{ahe,ahe2} and due to the SO interaction.\cite{chud}
We now consider an open circuit condition where the ensemble average of $\langle{\bm k}\pm|{\bm v}|{\bm k}\pm\rangle$ is zero, and focus on the effects of the spin electric field $\bm{{\cal E}}_\pm$.

The spin electric field (\ref{es}) accelerates the electrons with majority and minority spins in opposite directions to each other, giving rise to a spin current accompanied by an electric voltage.
The first term in Eq.~(\ref{es}) has been known as the origin of the conventional SMF.\cite{barnes,volovik,sanvito,yamane}
The second term reflects the nonadiabaticity in the electron spin dynamics,\cite{duine,tserkov,shibata} which goes to zero in the adiabatic limit $J_{\rm ex}\tau_{\rm sf}\rightarrow\infty$.
Since these two terms depend on $\partial{\bm m}/\partial t$ and $\nabla{\bm m}$, the appearance of the SMFs due to these spin electric fields are spatially confined in time-varying and spatially nonuniform magnetization regions.
The last two terms in Eq.~(\ref{es}), which contain the SO parameter $\eta_{\rm so}$, do not depend on $\nabla{\bm m}$. 
The fourth term, reflecting the nonadiabatic dynamics of the electron spin, was recently derived in the Rashba SO coupling systems based on the diagrammatic calculation.\cite{tatara}
In Ref.~\onlinecite{lee}, the spin electric field proportional to $\partial{\bm m}/\partial t\times{\bm E}$ was found in the Rashba SO coupling systems, where the electric field ${\bm E}$ due to the inversion asymmetry is assumed to be static.
We found that this spin electric field is just a part of a spin electric field proportional to $\partial({\bm m}\times{\bm E})/\partial t$, i.e., the third term in Eq.~(\ref{es});
there appears an additional spin electric field proportional to ${\bm m}\times\partial{\bm E}/\partial t$.

It should be noted that since the new spin electric field can be induced with static and uniform magnetization, one can investigate the SMF electrically in detail under no disturbance arising from the inductive voltage, in contrast to the other SMFs.
In addition, this term may be tuned via electric fields with variable frequencies, whereas the time depedence of the other terms are governed by the magnetization dynamics.
In the following, we propose two systems to demonstrate the SMF.

\section{Voltage measurement}
Let us consider a thin film of ferromagnetic conductor, which has a static and uniform magnetic structure ${\bm m}({\bm r},t)=\hat{\bm{y}}$ and is subjected to a space-independent ac electric field ${\bm E}({\bm r},t)=E_0 \sin\omega t\hat{\bm{z}}$, with $E_0$ and $\omega$ the amplitude and the angular frequency of the electric field, respectively, and $\hat{\bm{i}}$ the unit vector along the $i$ axis ($i=x$, $y$, or $z$).
Here $\hat{\bm{z}}$ axis is set to be normal to the film plane [see Fig.~1(a)].
In this condition, Eq.~(\ref{es}) is reduced to
\begin{equation}
\bm{{\cal E}}_\pm = \pm\frac{m_{\rm e}\eta_{\rm SO}}{\hbar} E_0\omega\cos\omega t \hat{\bm{x}}.
\label{es2}\end{equation}
Our sample may be as thin as the order of the Thomas-Fermi screening length, i.e., consisting of one or two atomic layers, so that external electric fields can survive to some extent inside the film though it is conducting.
In another case, applying gate voltages allows one to control the intrinsic electric fields due to the inversion asymmetry in semiconductors.\cite{soi1,soi2}
In this section, the electric voltage due to the spin electric field (\ref{es2}) is investigated.

\begin{figure}[t]
    \begin{center}
        \begin{tabular}{cc}
            \resizebox{80mm}{!}{\includegraphics{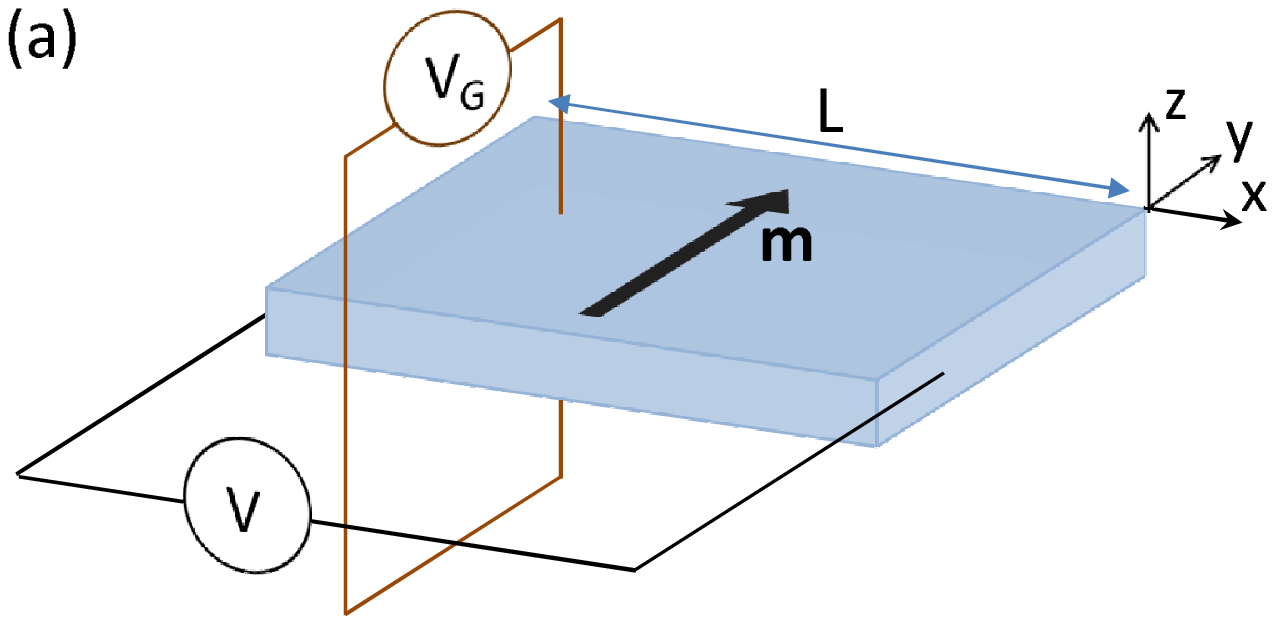}}\\
            \resizebox{80mm}{!}{\includegraphics{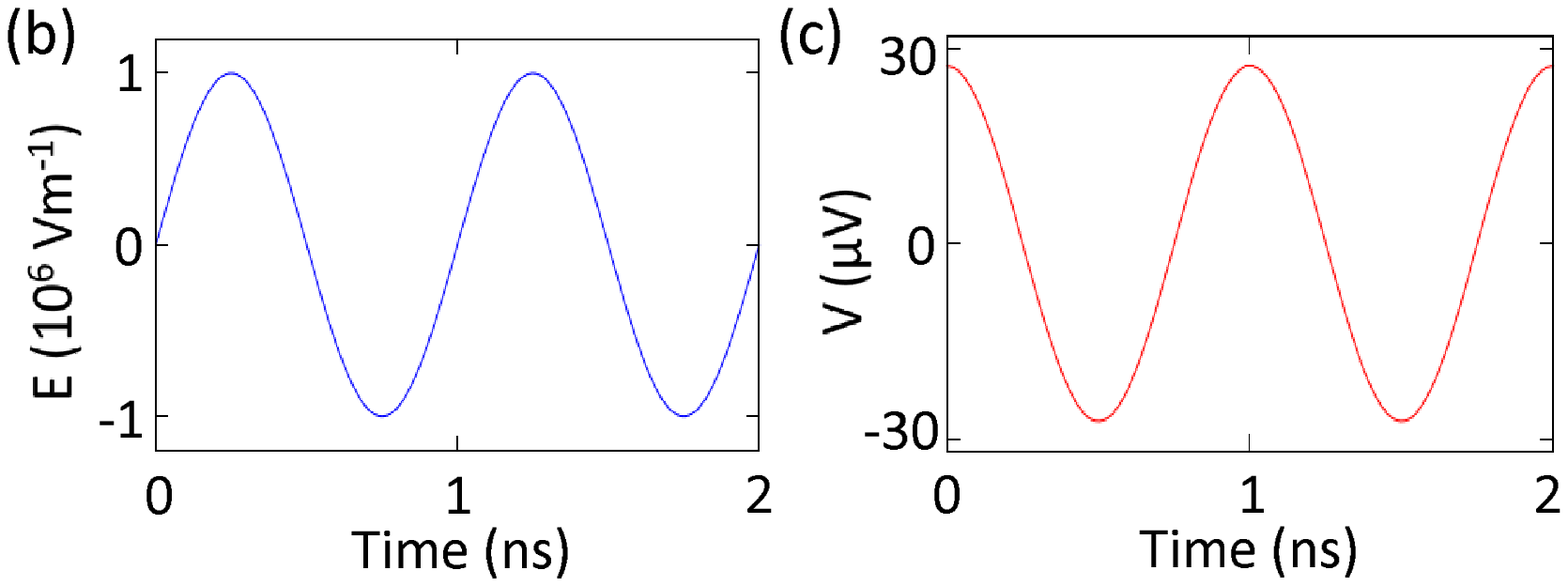}}\\
        \end{tabular}
    \caption{  (a) Measurement scheme of the electric voltage $V$ due to the SMF with the static and uniform magnetization ${\bm m}$.
                   The electric field varying sinusoidally in time is applied by the gate voltage $V_G$.
                   The electric  voltage appears perpendicular to both the time derivative of the electric field and the magnetization.
               (b), (c) Time evolution of a certain applied electric field and the corresponding electric voltage $V$.
                   The ac electric field gives rise to the ac electric voltage with the same frequency and the $\pi/2$ phase shift.
                   The amplitude of $V$ is proportional to the angular frequency of the electric field and the distance between the electrodes $L$.
            }
        \label{fig1}
    \end{center}
\end{figure}

The difference in the electric conductivities of the majority and minority electrons, $\sigma_F^\uparrow$ and $\sigma_F^\downarrow$, respectively, results in the charge current ${\bm j}_{\rm c}(t)$, which is given by
\begin{equation}
{\bm j}_{\rm c}(t) = \sigma_F^\uparrow \bm{{\cal E}}_+ + \sigma_F^\downarrow\bm{{\cal E}}_- = \frac{P\sigma_F m_{\rm e}\eta_{\rm SO}}{\hbar}E_0\omega\cos\omega t\hat{\bm{x}}.
\label{jc}\end{equation}
Here $P$ is the spin polarization defined as $P=(\sigma_F^\uparrow-\sigma_F^\downarrow)/(\sigma_F^\uparrow+\sigma_F^\downarrow)$ and $\sigma_F=\sigma^\uparrow_F+\sigma_F^\downarrow$.
The complex admittances should be used instead of the conductivities as we are considering the ac charge current.
However, for simplicity here we consider a condition where the reactance of the circuit is small enough so that the admittances are well approximated by the conductivities.

In the open circuit condition, the charge current (\ref{jc}) is canceled by the electric charge rearrangement, giving rise to an electric potential distribution $\phi(x,t)$ so that ${\bm j}_c(t)-\sigma_F\partial\phi(x,t)/\partial x=0$.
The electric voltage $V$ appearing between the sample edges, where $x=-L$ and $0$, is provided by
\begin{equation}
V=\int_{-L}^0 dx \frac{\partial\phi(x,t)}{\partial x} = \frac{P m_{\rm e}\eta_{\rm SO}L}{\hbar}E_0\omega\cos\omega t.
\label{v}\end{equation}
The amplitude of the voltage can be tuned by the distance between the electrodes $L$ and the angular frequency of the electric field $\omega$.
Notice that $V$ and ${\bm E}$ vary in time with the same angular frequency $\omega$, but their phases are different by $\pi/2$ since the spin electric field is proportional to the time derivative of ${\bm E}$;
$V\propto\cos\omega t$, while $E_z\propto\sin\omega t$, indicating that one can readily distinguish the SMF signal from the possible anomalous Hall voltage, which is proportional to ${\bm E}$ itself.
No inductive voltage appears in the present system because there is no dynamical magnetization.

In Figs.~1(b) and 1(c), the time evolution of Eq.~(\ref{v}) is shown together with that of the electric field.
The amplitude of $V$ is $\sim$30 $\mu$V, adopting the typical parameters in a thin film of ferromagnetic metals:
$m_{\rm e} = 9.1\times 10^{-31}$ kg, $\eta_\mathrm{so} = 10^{-21}$ m$^2$, $P=0.5$, $E_0=10^6$ V/m, $\omega=2\pi\times 10^9$ s$^{-1}$, and $L=1$ mm.

\begin{figure}[t]
    \begin{center}
        \begin{tabular}{cc}
            \resizebox{80mm}{!}{\includegraphics{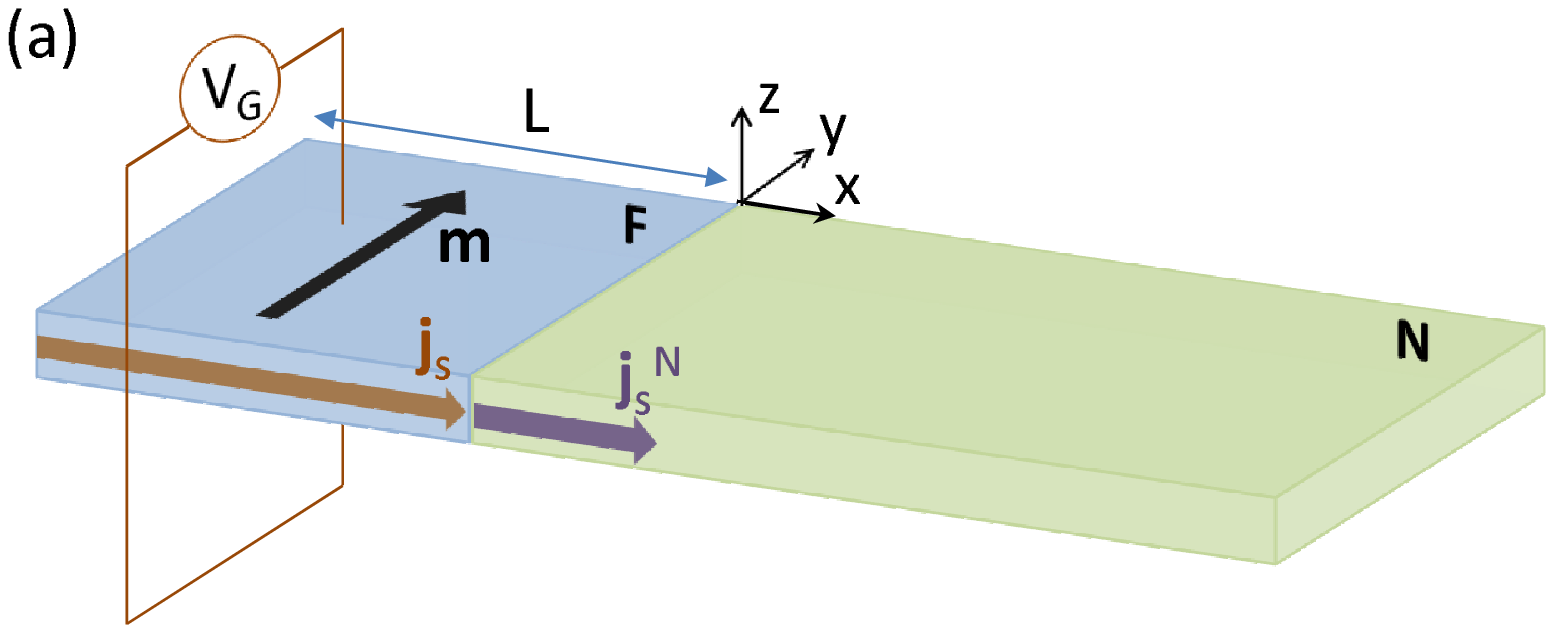}}\\
            \resizebox{80mm}{!}{\includegraphics{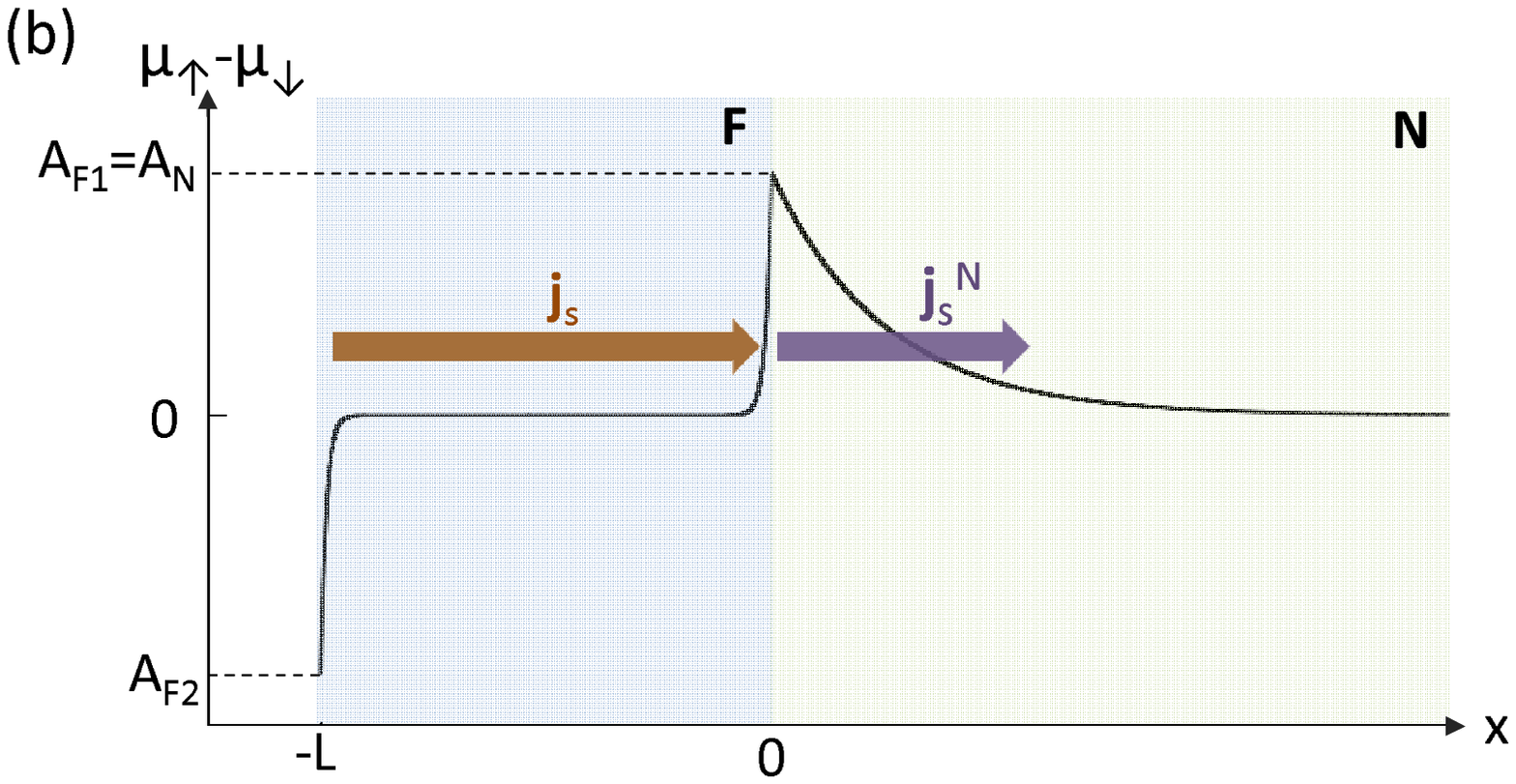}}\\
        \end{tabular}
    \caption{  (a) Schematic of the spin injection from a ferromagnet (F) into a nonmagnet (N).
                   The spin current ${\bm j}_{\rm s}$ induced in F due to the spin electric field (\ref{es2}) gives rise to a spin accumulation at the F/N interface, which decays in N as the diffusive spin current ${\bm j}_{\rm s}^N$ within the spin diffusion length.
               (b) Spatial dependence of the spin accumulation.
                   The spatial gradient of the spin accumulation gives the injected spin current ${\bm j}_{\rm s}^N$.
                   The coefficients $A_{F_1}$, $A_{F_2}$ and $A_N$ are determined by solving the diffusion equation (\ref{vf}) for F and N layers.
            }
        \label{fig2}
    \end{center}
\end{figure}

\section{Spin Injection}
Next, we investigate a spin injection method by using the spin electric field (\ref{es2}).
Let us consider a nonmagnetic conductor (N) attached to the ferromagnet (F), which has the in-plane magnetization and is subjected to the sinusoidally varying electric field as before [Fig.~2(a)].
In F layer, the spin electric field (\ref{es2}) induces not only the charge current (\ref{jc}) but also the spin current
\begin{equation}
{\bm j}_{\rm s} = - \left(\sigma_F^\uparrow\bm{{\cal E}}_+ - \sigma_F^\downarrow\bm{{\cal E}}_- \right)
 = - \frac{\sigma_F m_{\rm e}\eta_{\rm SO}}{\hbar}E_0\omega\cos\omega t \hat{{\bm x}},
\label{js}\end{equation}
giving rise to a spin accumulation at the ends of F, which diffuses into N and decays within the spin diffusion length [Fig.~2(b)].
The injected spin current into N, ${\bm j}_{\rm s}^N$, is calculated below.

The spin accumulations in F and N, $\mu_{F(N)}^\uparrow - \mu_{F(N)}^\downarrow$, with $\mu_{F(N)}^{\uparrow(\downarrow)}$ the electrochemical potential for a electron with majority (minority) spin in F (N), obeys the diffusion equation\cite{spincurrent,tserkov}
\begin{equation}
\nabla^2 ( \mu_{F(N)}^\uparrow - \mu_{F(N)}^\downarrow ) = \frac{1}{\lambda_{F(N)}^2} ( \mu_{F(N)}^\uparrow - \mu_{F(N)}^\downarrow ) - 2 e \nabla \cdot \bm{\mathcal{E}}_+,
\label{vf}\end{equation}
where $\lambda_{F(N)}$ is the spin diffusion length in F (N).
By substituting the spin electric field (\ref{es2}), which appears only in F, into Eq.~(\ref{vf}), the forms of the solutions are
\begin{equation}
\mu_F^\uparrow - \mu_F^\downarrow = A_{F1} e^{x/\lambda_F} - A_{F2} e^{\left(x+L\right)/\lambda_F},
\label{muf}\end{equation}
\begin{equation}
\mu_N^\uparrow - \mu_N^\downarrow = A_N e^{-x/\lambda_N},
\label{mun}\end{equation}
where the origin of the $x$ axis is located at the F/N interface.
Here we assume that N is much wider than $\lambda_N$ in the $x$ direction, so that the spin accumulation at another end of N can be neglected.
The coefficients $A_{F1}$, $A_{F2}$, and $A_N$ are determined from the boundary conditions for the electrochemical potentials, $\mu_F^{\uparrow(\downarrow)}(0,t)=\mu_N^{\uparrow(\downarrow)}(0,t)$, the spin current, ${\bm j}_{\rm s}(-L,t)=0$ and ${\bm j}_{\rm s}(0,t)={\bm j}_{\rm s}^N(0,t)$, and the charge current that is zero both in F and N because of the open circuit condition.
Thus we obtain
\begin{equation}
A_{F1} = A_N =  - \frac{1}{1+\alpha} 2 e \lambda_F  \frac{m_{\rm e}\eta_\mathrm{so}}{\hbar}E_0\omega \cos\omega t,
\end{equation}
\begin{equation}
A_{F2} =  2 e \lambda_F \frac{m_{\rm e}\eta_\mathrm{so}}{\hbar}E_0\omega\cos\omega t,
\end{equation}
where $\alpha$ is a dimensionless parameter defined as
\begin{equation}
\alpha = \frac{\lambda_F \sigma_N }{ \lambda_N \sigma_F (1-P^2)},
\label{a}\end{equation}
with $\sigma_N$ the electric conductivity of N.
The spin current in N is given by
\begin{equation}
{\bm j}_{\rm s}^N = - \frac{\sigma^N}{e} \nabla \left( \mu_N^\uparrow - \mu_N^\downarrow \right),
\label{jsn}\end{equation}
which oscillates in time with the angular frequency $\omega$.
Adopting the same parameters as before and $\lambda_N=1$ $\mu$m, $\lambda_F=10$ nm, and $\sigma_F = \sigma_N=$ (1 $\mu\Omega{\rm cm})^{-1}$, the amplitude of ${\bm j}_{\rm s}(0,t)$ at the F/N interface is $\sim$$5\times 10^4$ A/m$^2$.


\section{Conclusion}
In conclusion, the theory of spinmotive force has been extended in a system with spin-orbit coupling, and a new spinmotive force was derived, which can be induced by time-varying electric fields with static and uniform magnetization.
This spinmotive force has two advantages compared with the other SMFs.
(i) The electrical measurement of the spinmotive force is free from the inductive voltage.
(ii) The spinmotive force can be tuned by the electric fields free from the characteristic frequencies inherent in ferromagnets such as the ferromagnetic resonance frequency.
We have demonstrated the spinmotive force in two systems:
electric voltage measurement in a single ferromagnet and spin injection from a ferromagnet into a nonmagnetic conductor.


\section*{Acknowledgments}
The authors would like to thank Jairo Sinova and Jacob Gyles of Texas A\&M University, Makoto Kohda of Tohoku University, and Stewart E. Barnes of University of Miami for valuable discussions and comments.
This research was supported by the grant from a Grant-in-Aid for Scientific Research from MEXT, Japan, and Research Fellowship for Young Scientists from Japan Society for the Promotion of Science.


\end{document}